# Merger rates in hierarchical models of galaxy formation. II. Comparison with N-body simulations

Cedric Lacey[1,3] and Shaun Cole[2,4]
[1] *Physics Department, Oxford University, Nuclear & Astrophysics Building, Keble Rd, Oxford OX1 3RH*
[2] *Department of Physics, University of Durham, Science Laboratories, South Rd, Durham DH1 3LE*
[3] *cgl@thphys.ox.ac.uk*
[4] *Shaun.Cole@durham.ac.uk*

24 May 1994

**ABSTRACT**

We have made a detailed comparison of the results of N-body simulations with the analytical description of the merging histories of dark matter halos presented in Lacey & Cole (1993), which is based on an extension of the Press-Schechter method (Bond *et al.* 1991; Bower 1991). We find the analytical predictions for the halo mass function, merger rates and formation times to be remarkably accurate. The N-body simulations used $128^3$ particles and were of self-similar clustering, with $\Omega = 1$ and initial power spectra $P(k) \propto k^n$, with spectral indices $n = -2, -1, 0$. The analytical model is however expected to apply for abitrary $\Omega$ and more general power spectra. Dark matter halos were identified in the simulations using two different methods and at a range of overdensities. For halos selected at mean overdensities $\sim 100 - 200$, the analytical mass function was found to provide a good fit to the simulations with a collapse threshold close to that predicted by the spherical collapse model, with a typical error of $\lesssim 30\%$ over a range of $10^3$ in mass, which is the full dynamic range of our N-body simulations. This was insensitive to the type of filtering used. Over a range of $10^2 - 10^3$ in mass, there was also good agreement with the analytical predictions for merger rates, including their dependence on the masses of the two halos involved and the time interval being considered, and for formation times, including the dependence on halo mass and formation epoch.

The analytical Press-Schechter mass function and its extension to halo lifetimes and merger rates thus provide a very useful description of the growth of dark matter halos through hierarchical clustering, and should provide a valuable tool in studies of the formation and evolution of galaxies and galaxy clusters.

## 1 INTRODUCTION

In the standard cosmological picture, the mass density of the universe is dominated by collisionless dark matter, and structure in this component forms by hierarchical gravitational clustering starting from low amplitude seed fluctuations, with smaller objects collapsing first, and then merging together to form larger and larger objects. The halos of dark matter formed in this way, objects which are in approximate dynamical equilibrium, form the gravitational potential wells in which gas collects and forms stars to produce visible galaxies. Subsequent merging of the dark halos leads to formation of groups and clusters of galaxies bound together by a common dark halo, and is accompanied by some merging of the visible galaxies with each other. It is obviously of great interest to understand this process of structure formation via merging in more detail. One approach, begun by Aarseth, Gott & Turner (1979) and Efstathiou & Eastwood (1981), is to calculate the non-linear evolution of the dark matter numerically, using large N-body simulations. A second approach, complementary to the first, is to develop approximate analytical descriptions which relate non-linear properties such as the mass distribution and merging probabilities of collapsed objects to the initial spectrum of linear density fluctuations from which they grew. These analytical descriptions must then be tested against the results of the numerical simulations. If they work, they provide insight into the numerical results, and provide the basis for simplified calculations and modelling which can cover a much wider range of parameter space than is feasible with the numerical simulations alone.

The analytical approach was pioneered by Press & Schechter (1974) (hereafter PS), who derived, rather heuristically, an expression for the mass spectrum of collapsed, virialized objects resulting, via dissipationless gravitational clustering of initially cold matter, from initial density fluctuations obeying Gaussian statistics. The basis of the method is to derive a threshold value of the linear overdensity for collapse of spherical perturbations, and then calculate the fraction of mass in the linear density field that is above this threshold when smoothed on various scales. The PS mass function formula has since been widely applied to a variety of problems, including gravitational lensing by dark halos (Narayan & White 1988), abundance of clusters and their influence on the cosmic microwave background via the Sunyaev-Zel'dovich effect (Cole & Kaiser 1988), and galaxy formation (Cole & Kaiser 1989; White & Frenk 1991). An alternative derivation of the PS mass function was presented by Bond *et al.* (1991) (hereafter BCEK), who, by considering the random walk of linear overdensity (at a fixed location) as a function of smoothing scale, obtained a rigorous solution to the problem of above-threshold regions lying inside other above-threshold regions (the so-called "cloud-in-cloud" problem). BCEK also showed how the derived mass function depended on the filter used to define the spatial smoothing, and that the standard PS formula in fact only results in the case of "sharp $k$-space" filtering, which is also the only case for which exact analytical results can be obtained. (Related approximate results were also obtained by Peacock & Heavens (1990).) In addition, BCEK showed how to go beyond a calculation of the mass function at a single time, to derive the conditional mass function relating the halo masses at two different times. Indepedently, Bower (1991) extended the original method used by Press and Schechter and derived an identical expression for the conditional mass function. This extension to the PS theory was then taken up by Lacey & Cole (1993) (hereafter Paper I), who used the conditional mass function to derive a range of results on the merging of dark halos. These included the instantaneous merger rate as a function of the masses of both halos involved, and the distribution of formation and survival times of halos of a given mass identified at a given epoch, the formation time being defined as the earlier epoch when the halo mass was only half of that at the identification epoch, and the survival time as when the halo mass has grown to twice that at the identification epoch. The expressions for these depend on the initial linear fluctuations through their power spectrum, and on the background cosmology (density parameter $\Omega$ and cosmological constant $\Lambda$). Preliminary applications of these results were made to constraining the value of $\Omega$ from the merging of galaxy clusters, and to estimating the rate of accretion of satellites by the disks of spiral galaxies.

Kauffmann & White (1993) have extended the utility of the analytical results of BCEK and Bower (1991) by presenting a Monte Carlo method of generating merger trees, decribing the formation history of halos, that are consistent with the analytical conditional mass function. This technique has subsequently been utilized in studies of galaxy formation by Kauffmann, White & Guiderdoni (1993) and Kauffmann, Guiderdoni & White (1994). An alternative Monte Carlo implementation has also been used in studies of galaxy formation in Cole *et al.* (1994).

An alternative analytical approach is to assume that objects form from peaks in the initial density field. This has been extensively used to study clustering of galaxies and clusters (Peacock & Heavens 1985; Bardeen *et al.* 1986), but has been less used in calculating mass functions because of the problem of dealing properly with peaks which lie inside other peaks, and of identifying what mass object forms from a given size peak (*e.g.* Bond (1988)). Bond & Myers (1993a) have recently developed a new method which combines aspects of the Press-Schechter and peaks methods, but this requires one to generate and analyze Monte Carlo realizations of the linear density field, and so is much more complicated to apply, though still simpler than doing an N-body simulation.

Our aim in this paper is to test the analytical results for merging of dark halos derived in Paper I against a set of large N-body simulations. Specifically, we test the formulae for merger probabilities as a function of the masses of the halos involved and of the difference in cosmic epochs, and for the distribution of formation epochs of halos. We also revisit the question of how well the PS mass function itself works compared to simulations. The latter question has been investigated previously, notably by Efstathiou *et al.* (1988) (hereafter EFWD) for a set of self-similar models with power-law initial power spectra and $\Omega = 1$, and by Efstathiou & Rees (1988), White *et al.* (1993) and Bond & Myers (1993b) for Cold Dark Matter (CDM) models, who all found reasonably good agreement. BCEK tested the conditional mass function formulae against EFWD's simulations, and found encouraging results, but the size of their simulations ($32^3$ particles) did not allow very detailed comparisons. In this paper, we address these questions using $128^3 \approx 2 \times 10^6$ particle N-body simulations of self-similar models, with $\Omega = 1$ and initial power spectra $P(k) \propto k^n$, $n = -2, -1, 0$. With this large number of particles, it is possible to make fairly detailed tests of the merger formulae. There has been some discussion of what is the best filter to use with the standard PS mass function formula, and whether improved results can be obtained by choosing a threshold linear overdensity different from that of the simple spherical collapse model (Efstathiou & Rees 1988; Carlberg & Couchman 1989), and we investigate this also. Kauffmann & White (1993) made a qualitative comparison of some of the properties of the merger histories from their Monte Carlo method with those from a CDM simulation, and argued that there was reasonable agreement.

An important question that arises when comparing analytical predictions for mass functions, merger rates etc. of dark halos with the corresponding quantities in N-body simulations, is how best to identify the halos in the simulations. Given that the halos found in simulations are neither completely isolated nor exactly in virial equilibrium, there seems no unique way to do this. Much work has been based on the percolation or "friends-of-friends" algorithm (Davis *et al.* 1985) (DEFW), in which particles are linked together with other particles into a group if the distance to the nearest group member is less than a certain fraction (usually taken to be 0.2) of the mean interparticle separation. However, other group-finding schemes have also been used (*e.g.* Warren *et al.* (1992), Bond & Myers (1993b)). It is important to know how much the comparison with analytical results depends on the group-finding scheme employed, so we will investigate the effect of varying the linking length in the friends-of-friends scheme, and also use an alternative method based on finding spheres of particles of a certain overdensity.

The plan of the paper is as follows: In §2 we review the analytical results on merging derived in Paper I. In §3 we describe the N-body simulations, and in §4 we describe our group-finding schemes. In §5 we compare the N-body mass functions with the PS formula. The following two sections test the predictions for merging against the simulations: merger probabilities as a function of mass in §6, and formation epochs in §7. We present our conclusions in §8.

## 2 REVIEW OF ANALYTICAL MERGER RESULTS

### 2.1 Spatial Filtering and Random Trajectories

In this section, we review the analytical results on merging of dark halos derived in Paper I, BCEK and Bower (1991). Central to the approach is *spatial filtering* of the *linear density field*. The initial conditions for structure formation are specified as a Gaussian random density field $\delta(x) = \rho(\mathbf{x})/\overline{\rho} - 1$ having some power spectrum $P(k)$, where $\overline{\rho}$ is the mean density. This field is smoothed by convolving it with spherically-symmetric filters $W_R(r)$ of various radii $R$, having Fourier representations $\widehat{W}_R(k)$:

$$\widehat{W}_R(k) = \int W_R(|\mathbf{x}|) \exp(-i\mathbf{k}.\mathbf{x}) \, d^3\mathbf{x}. \qquad (2.1)$$

The variance of the field after smoothing with a filter is related to the power spectrum by

$$\sigma^2(R) \equiv \langle \delta^2 \rangle_R = \int_0^\infty \widehat{W}_R^2(k) P(k) \, 4\pi k^2 \, dk \qquad (2.2)$$

We list below the filters that we will be using in this paper, and their Fourier representations. We also give the "natural volume" $V_f$ that is associated with a filter of radius $R_f$, defined to be the integral of $W_R(r)/W_R(0)$ over all space.

*Top Hat (TH):*

$$W_R(r) = \begin{cases} 3/\left(4\pi R_T^3\right) & r < R_T \\ 0 & r > R_T \end{cases} \qquad (2.3)$$

$$\widehat{W}_R(k) = \frac{3}{(kR_T)^3} \left[\sin(kR_T) - (kR_T)\cos(kR_T)\right] \qquad (2.4)$$

$$V_T = (4\pi/3) R_T^3 \qquad (2.5)$$

*Gaussian (G):*

$$W_R(r) = \frac{1}{(2\pi)^{3/2} R_G^3} \exp\left(-\frac{r^2}{2R_G^2}\right) \qquad (2.6)$$

$$\widehat{W}_R(k) = \exp\left(-\frac{k^2}{2R_G^2}\right) \qquad (2.7)$$

$$V_G = (2\pi)^{3/2} R_G^3 \qquad (2.8)$$

*Sharp k-space (SK):*

$$W_R(r) = \frac{1}{2\pi^2 r^3} \left[\sin(r/R_S) - (r/R_S)\cos(r/R_S)\right] \qquad (2.9)$$

$$\widehat{W}_R(k) = \begin{cases} 1 & k < 1/R_S \\ 0 & k > 1/R_S \end{cases} \qquad (2.10)$$

$$V_S = 6\pi^2 R_S^3 \qquad (2.11)$$

The filters are normalized according to the condition $\int_0^\infty W_R(r) 4\pi r^2 \, dr = \widehat{W}_R(k=0) = 1$. The natural volume of a filter is thus $V_f = 1/W_R(r=0)$. (In the case of the sharp $k$-space filter, the volume integrals are a little ill-defined if done in real space, since the integral $\int_0^r W_R(r) 4\pi r^2 \, dr$ actually oscillates around 1 as $r \to \infty$. If desired, this minor problem can be cured by multiplying $W_R(r)$ by $\exp(-\alpha r)$ before doing the integral, and taking the limit $\alpha \to 0$ afterwards.) The natural mass under a filter is then defined as $M_f = \overline{\rho} V_f$.

When the density fluctuations are small ($\delta \ll 1$), they grow according to linear perturbation theory, $\delta(\mathbf{x}, t) \propto D(t)$, where the linear growth factor $D(t)$ depends on the background cosmological model; for $\Omega = 1$, $D(t) \propto a(t) \propto t^{2/3}$, $a(t)$ being the cosmic expansion factor. (We are assuming that only the growing mode of linear perturbation theory is present.) The non-linear evolution can be calculated analytically for spherical perturbations (*e.g.* Peebles (1980); Paper I); for a uniform overdense spherical fluctuation, the collapse time depends on its initial linear overdensity. It is convenient to work in terms of the initial density field extrapolated according to linear theory to some fixed reference epoch $t_0$, for which we also take $a(t_0) = 1$; from now on, this is what we will mean by $\delta(\mathbf{x})$. In terms of this extrapolated $\delta$, a spherical perturbation of mean overdensity $\delta$ collapses at time $t$ if $\delta = \delta_c(t)$, where, for $\Omega = 1$, we have the usual result

$$\begin{aligned} \delta_c(t) &= \delta_{c0}/a(t) = \delta_{c0} (t_0/t)^{2/3} \\ \delta_{c0} &= 3(12\pi)^{2/3}/20 \approx 1.69 \end{aligned} \qquad (2.12)$$

(The generalization of this for $\Omega < 1$ is derived in Paper I; note that the second line of equation (2.1) of that paper contains a typographical error.)

The analytical formulae we now present for the halo mass functions, conditional mass functions and lifetimes are expressed in terms of this threshold $\delta_c(t)$, and $\sigma(M)$, the variance of the smoothed linear density field as a function of the smoothing mass.

### 2.2 Mass Function

The fraction of mass in halos with mass $M$, at time $t$, per interval $dM$, originally derived by Press & Schechter (1974), is (Paper I equation (2.10))

$$\frac{df}{dM}(M, t) = \frac{\delta_c(t)}{(2\pi)^{1/2} \sigma^3(M)} \left| \frac{d\sigma^2(M)}{dM} \right| \exp\left[-\frac{\delta_c(t)^2}{2\sigma(M)^2}\right] \qquad (2.13)$$

$\sigma(M)$ is assumed to decline monotonically with increasing $M$. (To make the correspondence with the equations in Paper I, note that we there used the notation $S = \sigma^2(M)$, $\omega = \delta_c(t)$.) Thus, the comoving number density of halos of mass $M$ present at time $t$, per $dM$, is

$$\frac{dn}{dM}(M, t) =$$

$$\left(\frac{2}{\pi}\right)^{1/2} \frac{\overline{\rho}}{M^2} \frac{\delta_c(t)}{\sigma(M)} \left|\frac{d\ln\sigma}{d\ln M}\right| \exp\left[-\frac{\delta_c(t)^2}{2\sigma^2(M)}\right] \quad (2.14)$$

where $\overline{\rho}$ is now the mean density at the reference epoch $t_0$. By defining $\nu \equiv \delta_c(t)/\sigma(M)$, (2.13) can be rewritten as

$$\frac{df}{d\ln\nu} = \left(\frac{2}{\pi}\right)^{1/2} \nu \exp(-\nu^2/2) \quad (2.15)$$

which is independent of the form of the fluctuation spectrum.

### 2.3  Conditional Mass Function and Merger Probability

The conditional probability for a mass element to be part of a halo of mass $M_1$ at time $t_1$, given that it is part of a larger halo of mass $M_2 > M_1$ at a later time $t_2 > t_1$, is found by considering two different thresholds, $\delta_c(t_1)$ and $\delta_c(t_2)$. The result, derived somewhat differently by BCEK and Bower (1991), for the conditional mass fraction per interval $dM_1$ is (Paper I equation (2.15))

$$\frac{df}{dM_1}(M_1,t_1|M_2,t_2) =$$
$$\frac{(\delta_{c1} - \delta_{c2})}{(2\pi)^{1/2}(\sigma_1^2 - \sigma_2^2)^{3/2}} \left|\frac{d\sigma_1^2}{dM_1}\right| \exp\left[-\frac{(\delta_{c1} - \delta_{c2})^2}{2(\sigma_1^2 - \sigma_2^2)}\right] \quad (2.16)$$

where $\sigma_1 = \sigma(M_1)$, $\sigma_2 = \sigma(M_2)$, $\delta_{c1} = \delta_c(t_1)$ and $\delta_{c2} = \delta_c(t_2)$. The only assumption made here about the function $\delta_c(t)$ is that it monotonically decreases with increasing $t$. The reverse conditional probability, for $M_2$ given $M_1$, is (Paper I equation (2.16))

$$\frac{df}{dM_2}(M_2,t_2|M_1,t_1) =$$
$$\frac{1}{(2\pi)^{1/2}}\frac{\delta_{c2}(\delta_{c1}-\delta_{c2})}{\delta_{c1}}\left[\frac{\sigma_1^2}{\sigma_2^2(\sigma_1^2-\sigma_2^2)}\right]^{3/2}$$
$$\times \left|\frac{d\sigma_2^2}{dM_2}\right| \exp\left[-\frac{(\delta_{c2}\sigma_1^2 - \delta_{c1}\sigma_2^2)^2}{2\sigma_1^2\sigma_2^2(\sigma_1^2-\sigma_2^2)}\right] \quad (2.17)$$

This is obviously the same as the probability for a halo of mass $M_1$ at $t_1$ to be incorporated into a halo of mass $M_2 > M_1$ at $t_2 > t_1$. Thus, if we set $M_2 = M_1 + \Delta M$ and $t_2 = t_1 + \Delta t$ in the above formula, we get the probability for a halo to gain mass $\Delta M$ by merging in time $\Delta t$. Taking the limit $\Delta t \to 0$ then gives the instantaneous merger rate as a function of $M_1$ and $\Delta M$ (equation (2.18) of Paper I).

### 2.4  Formation Times

Suppose one identifies a halo of mass $M_2$ at time $t_2$. At an earlier time, one can identify the progenitors of this halo. We define the formation time of the halo identified at epoch $t_2$ as the earliest time $t_f < t_2$ at which it has a progenitor of mass $M_1$ at least half of $M_2$. We find the cumulative probability distribution for $t_f$ (equation (2.26) of Paper I)

$$P(t_f < t_1 | M_2, t_2) =$$
$$\int_{M_2/2}^{M_2} \frac{M_2}{M_1} \frac{df}{dM_1}(M_1, t_1|M_2, t_2)\, dM_1 \quad (2.18)$$

The differential probability distribution for $t_f$ is then given by

$$\frac{dp}{dt_f}(t_f|M_2, t_2) =$$
$$\int_{M_2/2}^{M_2} \frac{M_2}{M_1} \left|\frac{\partial}{\partial t_f}\left[\frac{df}{dM_1}(M_1, t_f|M_2, t_2)\right]\right| dM_1 \quad (2.19)$$

We noted in Paper I that the expression corresponding to equation (2.19) actually leads to a slight mathematical inconsistency in some cases: for power-law power spectra $P(k) \propto k^n$ with $n > 0$, the probability density for $t_f$ goes slightly negative for small $t_2 - t_f$. In Paper I, we also derived formation time distributions based on a Monte Carlo method, which do not have the problem of negative probability density. These distributions have similar shapes to the analytical ones, but with the mean shifted. We will see in §7 that the analytical distribution gives a remarkably good fit to the N-body results.

### 2.5  Self-Similar Models

The analytical results presented above make no special assumptions about the functions $\sigma(M)$ and $\delta_c(t)$, except that they are monotonic. However, in testing these results against simulations, we will focus on self-similar models, in which the density parameter $\Omega = 1$, so that there is no characteristic time in the expansion of the universe, and in which the initial density fluctuations have a scale-free spectrum, $P(k) \propto k^n$. In this case, the evolution of structure should be self-similar in time. This has some advantages, to be discussed in §3. From equation (2.2), we obtain $\sigma(M) \propto M^{-(n+3)/6}$, in general. For $\sigma$ to decline with increasing $M$, we require $n > -3$, which is just a condition for structure to grow hierarchically, with small objects collapsing first and then merging to form larger objects. For the top hat filter, the integral in equation (2.2) only converges for $n < 1$. We will be considering N-body models with $n = -2, -1, 0$.

For $\Omega = 1$, density fluctuations grow as $D(t) \propto a(t) \propto t^{2/3}$ in linear theory, so that the r.m.s. fluctuation on comoving scale $k^{-1}$ is roughly $\sqrt{4\pi k^3 P(k)}\, a(t)$. Thus we can define a characteristic non-linear wavenumber $k_\star(t)$ by

$$4\pi k_\star^3(t)\, P(k_\star(t))\, a(t)^2 = 1 \quad (2.20)$$

At time $t$, fluctuations are of order unity and are starting to collapse on a comoving lengthscale $\sim k_\star(t)^{-1}$. In a self-similar model, this should be the only characteristic lengthscale for structure.

In our analytical expressions for mass functions, merger probabilities etc, it is convenient to define a filter-dependent characteristic mass scale $M_\star(t)$ by

$$\sigma(M_\star(t)) = \delta_c(t) \quad (2.21)$$

where $\delta_c(t) = \delta_{c0}/a(t)$ for $\Omega = 1$. The mass $M_\star$ is related to the filter-independent quantity $k_\star$ by

$$M_\star(t) = \gamma_f \left(\frac{c_f(n)}{\delta_{c0}}\right)^{6/(n+3)} \frac{\overline{\rho}}{k_\star^3(t)} \quad (2.22)$$

where $\overline{\rho}$ is the mean density at the reference epoch when $a = 1$. $\gamma_f$ relates the volume of a filter to its radius through $V_{\rm f} \equiv \gamma_{\rm f} R_{\rm f}^3$ (c.f. §2.1), and $c_{\rm f}(n)$, which enters through the relation (2.2) between $\sigma(R)$ and $P(k)$, is defined by

$$c_{\rm f}^2(n) = \int_0^\infty \widehat{W}_{R=1}^2(k)\, k^{n+2}\, dk \qquad (2.23)$$

For the filters we are using, $c_f(-2) = 1.373, 0.941, 1.000$, $c_f(-1) = 1.500, 0.707, 0.707$ and $c_f(0) = 2.170, 0.666, 0.577$ for $TH, G, SK$ filters respectively. From equations (2.20) and (2.22), the characteristic mass grows as $M_\star(t) \propto a^{6/(n+3)}$. The mass function (2.13) can then be rewritten as

$$\begin{aligned}\frac{df}{d\ln M} &= \left(\frac{2}{\pi}\right)^{1/2} \left(\frac{n+3}{6}\right) \left(\frac{M}{M_\star}\right)^{(n+3)/6} \\ &\quad \times \exp\left[-\frac{1}{2}\left(\frac{M}{M_\star}\right)^{(n+3)/3}\right]\end{aligned} \qquad (2.24)$$

in which form the mass and time only appear in the combination $M/M_\star(t)$. Similarly, the merger probability (2.17) can be rewritten as a distribution for $M_2/M_1$ depending on $M_1/M_\star(a_1)$ and $a_2/a_1$, and the formation time distribution (2.19) can be rewritten as a distribution for $a_{\rm f}/a_2$ depending on $M_2/M_\star(a_2)$.

### 2.6 Choice of filtering and collapse threshold

The BCEK derivation of the PS mass function and of the conditional mass function is based on sharp $k$-space filtering. On the hand, the original, more heuristic, derivation by PS themselves assumed top hat filtering, as did Bower's derivation of the conditional mass function. Most applications of the PS formula have followed the latter approach and used the $\sigma(M)$ relation for top hat filtering (with $M = (4\pi/3)\overline{\rho} R_T^3$), but some have instead used $\sigma(M)$ for Gaussian filtering (with $M = (2\pi)^{3/2}\overline{\rho} R_G^3$) (*e.g.* Efstathiou & Rees (1988)). Even for a given choice of filter, one can obtain different $\sigma(M)$ relations simply by choosing a different mass-radius relation from the "natural" one discussed in §2.1; after all, it is not obvious how the mass of a collapsed object is related to the profile of the filter used to identify it. BCEK suggested calculating the filter mass for a general filter from $M = (4\pi/3)\overline{\rho} R_T^3$, where the "eqivalent top hat" radius $R_T$ is defined through the relation $\sigma(R) = \sigma_{TH}(R_T)$, on the grounds that the collapse threshold $\delta_{\rm c}(t)$ is also calculated for a top hat spherical perturbation. For power-law power spectra, this requires making $\gamma_{\rm f}$ in equation (2.22) a function of spectral index as well as filter type, while for a general $P(k)$ it must be a function of $R$. This procedure is equivalent to using the $\sigma(M)$ relation for top hat filtering in formulae like (2.13) and (2.17), even if these formulae are derived for sharp $k$-space filtering.

In this paper, we adopt an empirical approach to the choice of filter and $M(R)$ relation. We will compare the results of N-body simulations to the formulae using top hat, Gaussian and sharp $k$-space filtering for $\sigma(R)$. We assume a mass-radius relation $M = \gamma_{\rm f} \overline{\rho} R^3$, with $\gamma_{\rm f}$ a constant depending on the filter type but independent of the power-spectrum, but will consider values of $\gamma_{\rm f}$ different from the "natural" ones given in §2.1.

A related issue concerns the choice of collapse threshold, which for $\Omega = 1$ boils down to a choice for $\delta_{\rm c0}$ in equation (2.12). While the spherical collapse model gives an unambiguous answer, one can take the view that, since real collapses have non-top hat and non-spherical density profiles, $\delta_{\rm c0}$ should be regarded as a phenomenological parameter, chosen to give the best fit to N-body results (*e.g.* Bond & Myers (1993b)). Since the PS mass function and other formulae depend on $\gamma_{\rm f}$ and $\delta_{\rm c0}$ only through $M_\star$ (equation 2.22), there is a degeneracy between these two parameters for any given spectral index $n$, but this degeneracy is lifted as soon as one considers results for different $n$. The choice of $\delta_{\rm c0}$ and $\gamma_{\rm f}$ will be considered in relation to the simulations in §5.

## 3 N-BODY SIMULATIONS

The simulations were performed using the high resolution particle-particle-particle-mesh ($P^3M$) code of Efstathiou *et al.* (1985) (EDFW) with $128^3 \approx 2 \times 10^6$ particles. The long-range force was computed on a $256^3$ mesh, while the softening parameter for the short-range force was chosen to be $\eta = 0.2(L/256)$, where $L$ is the size of the (periodic) computational box. This corresponds to the interparticle force falling to half of its point-mass value at a separation $r \approx 0.4\eta \approx L/3200$. Initial positions and velocities were generated by displacing particles from a uniform $128^3$ grid according to the Zel'dovich approximation, assuming the linear power spectrum and Gaussian statistics. We ran one simulation for each of $n = -2, -1, 0$. For $n = -1$ and $n = 0$, the initial amplitude of the power spectrum was chosen to equal the white noise level at the Nyquist frequency of the particle grid; for $n = -2$, this choice was found to lead to large departures from self-similar behaviour in the derived mass functions and related quantities, so this simulation was instead started when the amplitude was smaller by a factor 0.4. We adopted the convention of normalizing the expansion factor $a$ to 1 when the variance of the linear theory power spectrum in a top-hat sphere of radius $L/32$ was 1. With this choice, the initial expansion factors were $a_i = 0.2, 0.15, 0.06$ respectively for $n = -2, -1, 0$. The initial r.m.s. 1D displacements of the particles were approximately $0.9, 0.5, 0.25$ in units of the particle grid spacing. The time integration was performed using the variable $p = a^\alpha$, with $\alpha = 2/(n+3)$, and a constant stepsize $\Delta p/p_i = \frac{1}{4}\alpha(256\eta/L)^{3/2} \approx 0.023\alpha$ (EDFW).

For each simulation we output the positions and velocities of all the particles at many epochs. The spacing of these outputs was chosen so that the characteristic mass, $M_\star$, increased by a factor $\sqrt{2}$ between each successive output, which corresponds to an increase of a factor $2^{(n+3)/12}$ in the expansion factor $a$. The final expansion factors for the simulations were determined basically by the computer resources available; for the later stages, the CPU time was dominated by the short-range force calculation by a large factor. The simulations were stopped at expansion factors $a = 1.26, 2.00, 1.68$ respectively for $n = -2, -1, 0$. This gave us between 20 and 24 useful outputs from each simulation.

Self-similar models have two advantages from the point of view of testing our analytical predictions against simulations. (i) We can check whether the simulations obey the self-similar scaling which physically they should. In particular, this allows us to test whether the simulations were started at a small enough expansion factor. We can also delineate the range of masses of virialized objects for which the simulations are reliable, bearing in mind the effects of particle discreteness and force resolution on small scales, and missing long-wavelength modes on scales larger than the box. (ii) By using the self-similar scaling, we can straightforwardly combine the results from different output times to reduce the Poisson fluctuations on the N-body results due to the finite number of halos in the box. These factors, and the simple way to test for dependence on the form of the initial spectrum, were what motivated us to look at self-similar models rather than more physically-inspired models such as Cold Dark Matter.

## 4  GROUP-FINDING IN THE SIMULATIONS

### 4.1  Overview

We wish to obtain the properties of dark matter halos from the simulations to compare with our analytical results. Simple theoretical calculations idealize dark halos as isolated spherical objects in dynamical equilibrium, but the objects found in simulations with $\Omega = 1$ are neither isolated nor in complete dynamical equilibrium, because halos continue to grow by accreting or merging with other halos on a timescale comparable to the expansion time, which is also comparable to their internal dynamical times. (In a universe with $\Omega \ll 1$, on the other hand, one expects the conditions of isolation and dynamical equilibrium to be much better satisfied). Nor are real halos spherical. The issue of how to identify the groups of particles in the simulations that one calls dark halos is therefore not completely straightforward, and a variety of schemes have been used by different authors (*e.g.* DEFW, Barnes & Efstathiou (1987), Warren *et al.* (1992), Bond & Myers (1993b)). In order to give some idea of how sensitive the comparisons are to the group-finding method employed, we will present results for two different schemes, namely percolation, and a spherical overdensity method. Both are based on particle positions, making no use of the velocity information in the simulations. Both methods effectively involve choosing a density threshold to define groups (a local density in one case and a mean density inside a sphere in the other), but do not build in any preferred length or mass scales. The latter is important if one wishes to study properties like the distribution of halo masses. A fuller discussion and comparison of the internal properties of the halos found by using these different schemes and parameters, for the different initial power spectra, will be given by Cole & Lacey (1994).

### 4.2  Friends-of-Friends (FOF) Groups

The percolation method is the standard friends-of-friends algorithm (hereafter FOF) of DEFW, which has been widely used. Groups are defined by linking together all pairs of particles with separation less than $b\bar{n}^{-1/3}$, $\bar{n}$ being the mean particle density. This defines groups bounded approximately by a surface of constant local density, $\rho/\bar{\rho} \approx 3/(2\pi b^3)$. The value usually used for the dimensionless linking length is $b = 0.2$ (*e.g.* Frenk *et al.* (1988)), which corresponds to a density threshold $\rho/\bar{\rho} \approx 60$. For a spherical halo with a density profile $\rho(r) \propto r^{-2}$, this local density threshold corresponds to a mean overdensity $\langle\rho\rangle/\bar{\rho} \approx 180$, which is close to the value $18\pi^2 \approx 178$ for a just virialized object predicted for a top-hat spherical collapse (*e.g.* Peebles (1980), Paper I). It has been argued that the choice $b = 0.2$ approximately delineates between objects which are virialized and objects which are still collapsing in their outer parts, but in our own investigations (Cole & Lacey 1994), we find that the closeness to global virial equilibrium of N-body halos depends rather weakly on the value of $b$, so that this criterion does not strongly select a value for $b$. We will make comparisons with FOF groups for $b = 0.15, 0.2, 0.3$, corresponding to local overdensities of 140, 60 and 18 respectively. We will use the abbreviation FOF($b$) for groups identified using FOF with linking parameter $b$.

### 4.3  Spherical Overdensity (SO) Groups

The second method we apply, which we call spherical overdensity (SO), is based on finding spherical regions in a simulation having a certain mean overdensity, which we denote by $\kappa \equiv \langle\rho\rangle/\bar{\rho}$. We first calculate a local density for each particle by finding the distance $r_N$ to its N'th nearest neighbour, and define the density as $3(N+1)/(4\pi r_N^3)$. Particles are sorted by density. The highest density particle is taken as the candidate centre for the first sphere. A sphere is grown around this centre, with the radius being increased until the mean overdensity first falls below the value $\kappa$. (The sphere must contain at least 2 particles). The centre of mass of the particles in this sphere is then taken as a new centre, and the process of growing a sphere of the specified overdensity repeated. This process is iterated until the shift in the centre between successive iterations falls below $\epsilon \bar{n}^{-1/3}$. The particles in the sphere are all labeled as belonging to the same group, and removed from the list of particles considered by the group-finder. Then one moves on the next highest density particle which is not already in a group, and repeats the process of finding an overdense sphere, including iteration of the centre. Finally, after all the groups have been found, any groups which lie inside larger groups are merged with the larger group, according to the following procedure: Each group is considered as a sphere of mean overdensity $\kappa$, based on its actual mass, centred on its actual centre of mass. Starting with the largest sphere, any smaller sphere whose centre is inside the larger sphere is merged with that sphere (but the assumed radius of the larger sphere is not changed). This is then repeated for the next largest remaining sphere, and so on down in mass. We will use the abbreviation SO($\kappa$) for SO groups identified at overdensity $\kappa$.

We will present comparisons for SO groups with spherical overdensity $\kappa = 180$, chosen to agree with the spherical collapse model. For the local density estimate, we used $N = 10$, but the results were not especially sensitive to this. For the convergence of the group centres, we found $\epsilon = 0.1$ to be adequate. With these parameters, an average of fewer than 0.1 iterations per group were required, and the fraction of particles involved in merging small groups into large ones was less than $10^{-3}$. We also experimented with defining the initial list of centres for growing spheres from particles ranked either by gravitational potential (deepest potential first) or by the magnitude of the acceleration (highest acceleration first). The potential and acceleration were calcu-lated for each particle using a modified version of the $P^3M$ code, with the same grid and softening parameters as for the original simulation. The groups found starting from acceleration or potential centres were almost identical to those found starting from density centres, and the mass functions agreed to within a few percent. A fuller discussion of the method will be given in Cole & Lacey (1994). We have used the method based on density centres in this paper since it is more straightforward.

Our spherical overdensity algorithm has many similarities to the method used by Warren *et al.* (1992), and to the "smooth particle overdensity" method of Bond & Myers (1993b). The Warren *et al.* method grows spheres of specified overdensity around centres which are particle positions ranked by gravitational acceleration, and merges small halos which are inside larger ones, but does not iterate the positions of the centres. The Bond & Myers method grows spheres of a given overdensity around centres which are chosen initially to be positions of particles ranked by local density, and then iterates each sphere until the centre of mass converges. However, the volume of a group, used in defining its mean density, is defined by calculating a volume for each particle based on its local density, and summing these over all particles within the sphere to get the total volume, rather than as the volume of the sphere itself.

### 4.4 Comparison of Methods

The percolation method has the advantage that it is simple, relatively fast, and does not make any assumption about the geometry of the groups. However, some (a definite minority) of the groups it selects are formed of two or more dense lumps separated by low-density bridges, as has been noted by previous authors. These groups seem rather unphysical. The spherical overdensity method avoids this problem, instead producing groups concentrated around a single centre, but on the other hand, tends to chop off the outer portions of ellipsoidal halos. It is also more complicated and more time-consuming to run. It is not clear which method should be considered "best", so it is interesting to compare results obtained with both.

## 5 MASS FUNCTIONS IN THE SIMULATIONS

### 5.1 Tests for Self-Similarity

Figure 1 shows the N-body mass functions for all output times for all three values of the spectral index $n$. The halos were found using the friends-of-friends method with $b = 0.2$, and the mass function for each output time rescaled to a distribution in $M/M_\star(t)$, with $M_\star(t) \propto a(t)^{6/(n+3)}$ computed using a top hat filter with the standard choices $\gamma_{\rm f} = 4\pi/3$ and $\delta_{\rm c0} = 1.69$. Scaled in this way, the mass functions for different times should be identical, as a simple consequence of self-similarity, and it can be seen that the simulations conform to this expectation very well. We have excluded from the plots halos having $N < N_{\rm min} = 20$ or $N > N_{\rm max} = 2 \times 10^4$ particles. Halos containing only a few particles are not represented accurately by the simulations, and cause noticeable departures from self-similarity if included, while the properties of halos having masses approaching that of the box are affected by the absence from the initial conditions of modes with wavelength exceeding the size of the box. For our simulations, the $N_{\rm max}$ cutoff in fact makes little difference, because $M_\star$ is still much smaller than the mass of the box when the simulations are halted.

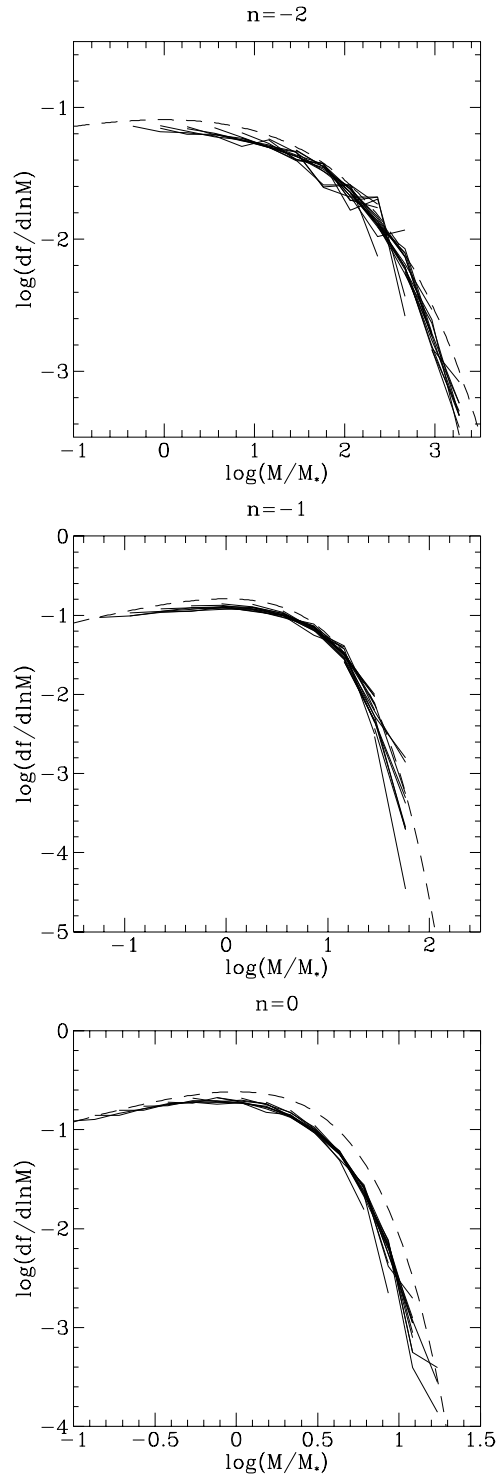

**Figure 1.** Mass functions for different output times, for groups identified using FOF(0.2), and masses rescaled to be in units of the characteristic mass $M_\star(t)$. $df/d\ln M$ is the fraction of mass in halos per logarithmic interval in halo mass. Each panel shows the results for a single N-body run ($n = -2, -1, 0$), with the mass functions for different output times plotted as solid lines, and the Press-Schechter prediction (equation (2.24)) for top-hat filtering and $\delta_{\rm c0} = 1.69$ as a dashed line.

Rigorously, self-similar clustering solutions only exist for spectral indices in the range $-1 < n < 1$, since outside this range, the r.m.s. peculiar velocity receives divergent contributions from either large or small scales (Davis & Peebles 1977). For $n < -1$, the problem arises from the very long wavelength fluctuations, which contribute negligibly to the r.m.s. overdensity, but produce large-scale coherent velocities. This should not affect the collapse of structure on small scales, so small scale structure is still expected to evolve in a self-similar way. This expectation is borne out by the scaling of the mass functions, which works as well for the $n = -2$ models as for $n = -1$ or $0$.

Since the scaling of the mass functions in the simulations is consistent with self-similarity, we average the results for all output times together to increase their precision, weighting the contribution to $df/d\ln M$ in each bin in $M/M_\star$ in proportion to the number of halos in that bin for each output time. We use these averaged mass functions in the comparisons that follow.

### 5.2 Choice of Filter and Density Threshold

We now consider the choice of filter and collapse threshold in the Press-Schechter mass function (*c.f.* §2.6). The PS prediction is that when the distribution in mass is converted to a distribution in $\nu = \delta_{\rm c}(t)/\sigma(M)$, it should have a universal form (equation 2.15), independent of the power spectrum. However, $\sigma(M)$, and thus $\nu$, depend on the filter used. For self-similar models, $\nu = (M/M_\star(t))^{(n+3)/6}$, and different choices of filter lead to different values of $M_\star$, according to equation (2.22). For a given filter, the value of $M_\star$ also depends on the parameters $\delta_{\rm c0}$ and $\gamma_{\rm f}$ (recall that the filter mass is related to the radius by $M_f = \gamma_{\rm f}\bar\rho R_f^3$).

We first test how well the N-body mass functions for different initial power spectra agree with each other when expressed as functions of $\nu$. Figure 2 shows the mass functions of FOF(0.2) groups in our self similar models, converted to distributions in $\nu$, for three different choices of filter: top hat (TH), Gaussian (G) or sharp $k$-space (SK). The *relative* placement of the N-body curves for different $n$ depends on $\gamma_{\rm f}$, but is independent of $\delta_{\rm c0}$, which just causes a uniform shift of all the curves by the same amount. For top hat and sharp $k$-space filtering we have used the "natural" values of $\gamma_{\rm f}$ from §2.1, while for Gaussian filtering, we have used both the natural value and one 2.5 times larger. We have assumed $\delta_{\rm c0} = 1.69$. It can be seen that for top hat and sharp $k$-space filtering, the N-body mass functions $df/d\ln\nu$ for different $n$ agree fairly well using the natural values for $\gamma_{\rm f}$ ($4\pi/3$ and $6\pi^2$ respectively), so we have not considered different values. For Gaussian filtering, on the other hand, there is a large spread between the N-body curves for different $n$ when $\gamma_{\rm f}$ is taken to be its natural value $(2\pi)^{3/2}$, but good agreement if the filter mass is increased by a factor 2.5. Previous authors who have used Gaussian filtering in conjuction with the PS formula (*e.g.* Efstathiou & Rees (1988), Carlberg & Couch-man (1989)) all appear to have assumed $\gamma_{\rm f} = (2\pi)^{3/2}$. The effect of the latter choice is that the best fitting value for $\delta_{\rm c0}$ depends on the shape of the power spectrum. These results about the best values for $\gamma_{\rm f}$ for different filters were found to apply equally well for the other group identification schemes we tested (friends-of-friends with $b = 0.15$ and $b = 0.3$, and spherical overdensity with $\kappa = 180$).

Having chosen optimal values for $\gamma_{\rm f}$, a second question is: how well does the PS formula actually fit the N-body results, and what is the best value to use for $\delta_{\rm c0}$? In Figure 2, which assumes $\delta_{\rm c0} = 1.69$, the standard PS prediction (equation 2.15) is shown by a dot-dash curve in each panel. The PS and N-body curves have very similar shapes, but it is apparent that they could be brought into even closer agreement by shifting the N-body curves horizontally, which corresponds to changing $\delta_{\rm c0}$ ($\nu \propto \delta_{\rm c0}$). For each filter, we have estimated by eye what value of $\delta_{\rm c0}$ gives the best fit of PS with N-body mass functions. The best fitting PS curve is shown by a dotted line in each panel, and the corresponding value of $\delta_{\rm c0}$ displayed. (Rather than replot the N-body curves with the new value of $\delta_{\rm c0}$, we have shifted the PS curve inversely as, $\nu \propto \delta_{\rm c0}^{-1}$.) For Gaussian filtering with the non-optimal $\gamma_{\rm f} = (2\pi)^{3/2}$, the best fit $\delta_{\rm c0}$ depends strongly on the power spectrum, so we have fitted the $n = -1$ results, in the middle of our range. Using the optimal $\gamma_{\rm f}$ values, it can be seen that the best fit $\delta_{\rm c0}$ values for FOF(0.2) groups differ by less than 20% from the canonical $\delta_{\rm c0} = 1.69$ for each of the three filters considered. When these best fit values are used, the PS formula fits the N-body mass functions extremely well, with an accuracy better than 30% over the range $0.3 \lesssim \nu \lesssim 3$. However, in all cases the PS formula systematically over-estimates the abundance of low mass halos ($\nu \lesssim 1$) compared to the simulations.

These results on $\delta_{\rm c0}$ are reasonably consistent with those found by previous authors. EFWD found a reasonable fit to FOF(0.2) groups in self-similar models for top hat filtering with $\delta_{\rm c0} = 1.68$, while Efstathiou & Rees (1988) and Carlberg & Couchman (1989) found $\delta_{\rm c0} = 1.33$ and $\delta_{\rm c0} = 1.44$ respectively for FOF(0.2) groups in CDM models using Gaussian filtering with $\gamma_{\rm f} = (2\pi)^{3/2}$ (note that the CDM power spectrum has effective spectral index $d\ln P(k)/d\ln k \approx -2$ on the range of scales considered). Bond & Myers (1993b) find $\delta_{\rm c0} = 1.58$ from their CDM simulations, but this is using their own group-finding scheme, which produces somewhat more massive groups than FOF(0.2), and correspondingly requires a smaller $\delta_{\rm c0}$.

### 5.3 Mass Functions with Different Group-finding

It is important to investigate how sensitive the results on mass functions are to the group-finding method employed. We have repeated the previous analysis for friends-of-friends with two other values of the linking parameter, $b = 0.15$ and $0.3$, and for the spherical overdensity method with overdensity $\kappa = 180$. Figure 3 shows the results for top hat filtering with $\gamma_{\rm f} = 4\pi/3$ and $\delta_{\rm c0} = 1.69$. The dot-dash curve in each panel is identical, and is the standard PS result, while the dotted curve is the shifted PS curve which seems to give the best fit, the corresponding best fit value of $\delta_{\rm c0}$ being given in each panel. Comparing the N-body mass functions with different group finders, the plots show the expected result that halo masses are smaller for FOF(0.15) than FOF(0.2), and larger for FOF(0.3). SO(180) halos are also on average smaller than FOF(0.2), but give very similar mass functions to FOF(0.15). The best fit values of $\delta_{c0}$ vary with group-finder accordingly, larger halo masses requiring a smaller $\delta_{c0}$, and vice versa, as shown in the figure. With TH filtering, all the group-finders give reasonable fits to PS (with different $\delta_{c0}$), though in each case, when the PS formula is fitted at the high mass end, it predicts somewhat too many low mass halos, the discrepancy being largest for FOF(0.15) and SO(180). FOF(0.3) groups result in the best fit to PS overall, when $\delta_{c0}$ is allowed to vary. For the canonical choice $\delta_{c0} = 1.69$, FOF(0.2) and FOF(0.3) groups agree about equally well with PS. For Gaussian and sharp $k$-space filtering, with $\gamma_f = 2.5\,(2\pi)^{3/2}$ and $6\pi^2$ respectively, the fits of the PS formula to the N-body mass functions for the different group finders are about as good as for top hat filtering when $\delta_{c0}$ is allowed to vary.

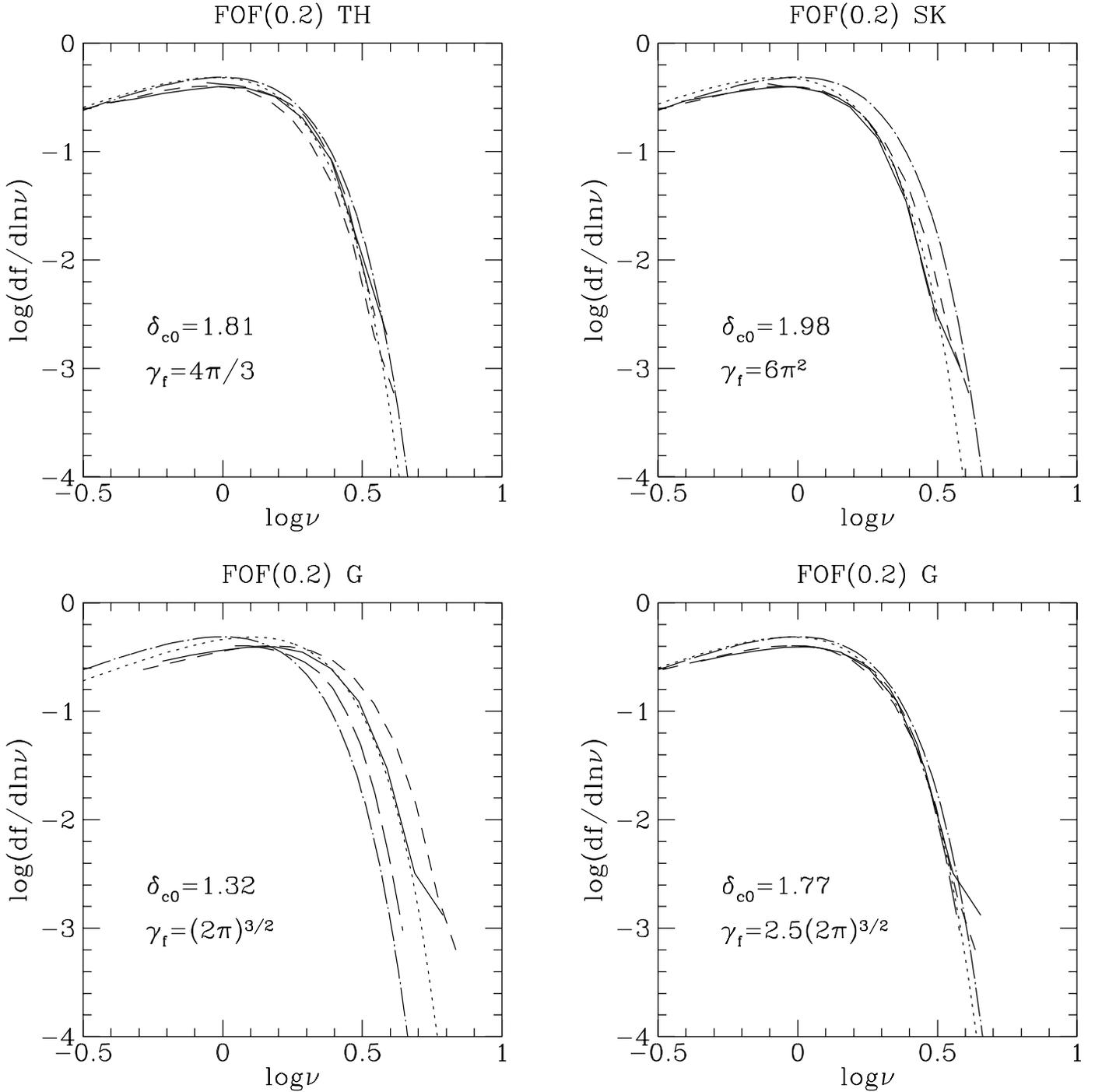

**Figure 2.** Averaged mass functions for the different N-body models, compared to the Press-Schechter mass function using different filters. $df/d\ln\nu$ is the fraction of mass per logarithmic interval in $\nu = (M/M_*)^{(n+3)/6}$. Groups were identified using FOF(0.2). Each panel shows results for a different choice of filter: top hat (TH), Gaussian (G) and sharp $k$-space (SK) respectively. For Gaussian filtering, we show results for two different values of $\gamma_f$, which relates the filter mass to filter radius through $M_f = \gamma_f \bar{\rho} R_f^3$. Each panel shows the N-body mass functions for $n = -2$ (long dashed line), $n = -1$ (solid) and $n = 0$ (short dashed), where these have been averaged over all output times. Also shown is the standard Press-Schechter result (equation (2.15)) (long dash dot curve), and the PS curve shifted in $\nu$ to give the best fit to the N-body results (dotted curve). Shifting in $\nu$ is equivalent to using a different value for $\delta_{c0}$ from the standard one. The values of $\gamma_f$ assumed, and of $\delta_{c0}$ for the best-fitting PS curves, are shown in the lower left corner of each plot.

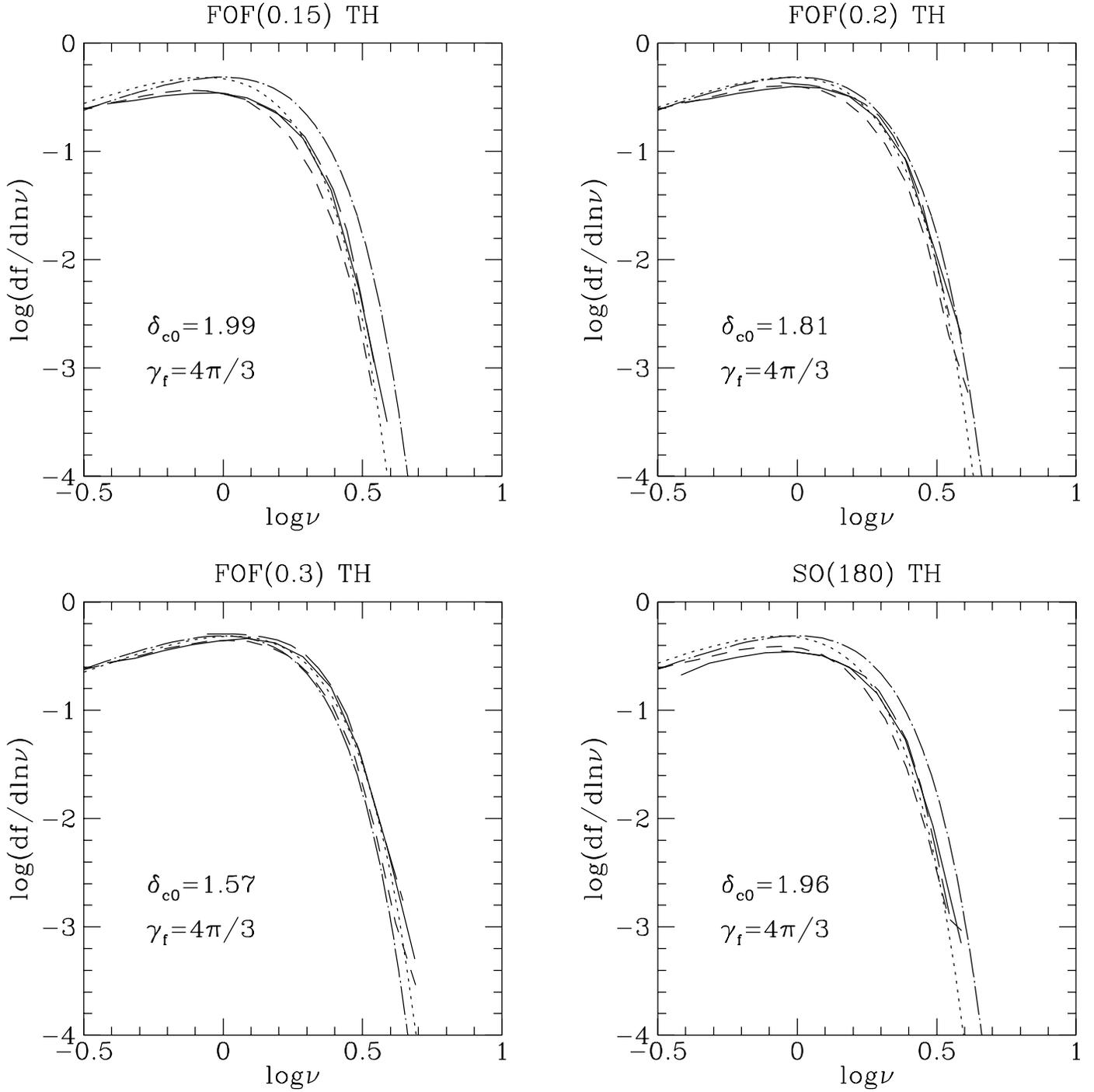

**Figure 3.** Effects on N-body mass functions of different group-finding schemes. Each panel shows the mass functions for $n = -2, -1, 0$ (long dashed, solid and short dashed lines respectively), for a different choice of group finding scheme and parameters. The first 3 panels show friends-of-friends with $b = 0.15, 0.2, 0.3$ respectively, while the last shows the spherical overdensity method with $\kappa = 180$. In each panel, the N-body mass functions have been averaged over output times, and converted to distributions in $\nu$ assuming top hat filtering with $\gamma_f = 4\pi/3$ and $\delta_{c0} = 1.69$. Also shown in each panel is the standard Press-Schechter result (equation (2.15)) (long dash dot curve), and the PS curve shifted in $\nu$ to give the best fit to the N-body results (dotted curve). The values of $\delta_{c0}$ corresponding to the best-fitting PS curves are shown in the lower left corner of each plot.

### 5.4 Discussion

To summarize the results of this section: at least for power-law initial power spectra, $P(k) \propto k^n$ with $-2 \leq n \leq 0$, there seems little to choose between top hat, Gaussian or sharp $k$-space filtering in the Press-Schechter mass function, provided the factor $\gamma_f$ is chosen appropriately. For top hat and sharp $k$-space filtering, the natural values ($4\pi/3$ and $6\pi^2$ respectively) work well, while for Gaussian filtering, a value around $2.5\,(2\pi)^{3/2}$ seems best. The N-body mass functions change with changes in the group-finding method or in its parameters, but this can to a considerable extent be compensated for in the PS formula by adjusting $\delta_{c0}$. If $\delta_{c0}$ is instead fixed at the value 1.69 from the spherical collapse model, then for FOF groups, a value $b \approx 0.2 - 0.3$ seems to give the best agreement with PS. We caution however that these conclusions may be changed for more general power spectra, in particular, there may be more difference between the top hat filter, which cuts off only as $\widehat{W}_R(k) \propto k^{-2}$ at short wavelengths, and the Gaussian or sharp $k$-space filters, which both cut off much more sharply. In the remainder of this paper, in looking at merger properties, we will concentrate on $b = 0.2$ FOF groups, using top hat filtering with $\gamma_f = 4\pi/3$, since these choices are the most standard. For simplicity we will use the canonical value $\delta_{c0} = 1.69$ for the collapse threshold, since this is in any case close to the best fit for the other choices made. The PS mass function then fits our N-body results to within a factor 2 or better for $0.3 \lesssim \nu \lesssim 3$.

## 6 MERGER RATES

The conditional mass function, $df/d\ln \Delta M |_{M_1, \Delta a, a_1}$, (2.17) describes the probability that a halo of mass $M_1$ at the epoch when the expansion factor equals $a_1$ will accrete mass $\Delta M \equiv M_2 - M_1$ to become a halo of mass $M_2$ at expansion factor $a_2 = a_1 + \Delta a$. In the limit of $\Delta a \to 0$ this yields the instantaneous merger rate at expansion factor $a_1$ of halos of mass $M_1$ with halos of mass $\Delta M$. Here we compare the conditional mass function (2.17) for both small and large time intervals $\Delta t$ with estimates made directly from our N-body simulations.

Having constructed group catalogues for each output time of our simulations using one of the group finding algorithms discussed in §4 it is an easy matter to construct the conditional mass function for any pair of output times. For each group of mass $M_1$ identified at the epoch when the expansion factor equals $a_1$ we determine which halo it has become incorporated in at expansion factor $a_2$ by finding the halo at this later epoch which contains the largest fraction of the particles from the original halo. Defining the mass of this new halo as $M_2$ we construct a joint histogram of $M_1$ and $\Delta M = M_2 - M_1$. The scale free nature of the initial conditions of our simulations implies that the form of these histograms when expressed in units of $M_1/M_\star$ ($M_\star$ given by (2.21) at $a = a_1$) and $\Delta M/M_1$ should depend only on $\Delta a/a_1 \equiv (a_2 - a_1)/a_1$. Consequently the histograms from different pairs of output times, but with the same value of $\Delta a/a_1$, can be averaged to yield more accurate estimates of the conditional mass function. For each pair of output times we weight the contribution to $df/d\ln \Delta M |_{M_1, \Delta a, a_1}$ in each bin of $\Delta M/M_1$ in proportion to the number of halos in that bin. We estimate Poisson error bars for the conditional mass function from the number of halos $N_{\rm bin}$ in each mass bin, summed over the pairs of output times. Successive pairs of output times are not really independent, so we define an effective number per bin as $N_{\rm eff} = N_{\rm bin}/f$, and take the fractional error to be $1/\sqrt{N_{\rm eff}}$. For the particular case for which $\Delta a/a_1$ corresponds to the spacing between successive outputs, *i.e.* outputs spaced by a factor $\sqrt{2}$ in $M_\star(t)$, we take $f = 1$, but for all larger values of $\Delta a/a_1$ we adopt $f = 2$, which corresponds to taking outputs separated by a factor 2 in $M_\star(t)$ to be independent. This procedure is obviously not rigorous, but provides some indication of the magnitude of statistical errors.

We have compared the analytical predictions of (2.17) with the conditional mass functions estimated from each of our the simulations for a wide range of both $M_1$ and $\Delta a/a_1$. A representative selection of these comparisons are shown in Figures 4, 5 and 6, which are for the case of groups identified using FOF(0.2), and restricted to halos satisfying $N_1 > 20$ and $\Delta N = N_2 - N_1 > 20$, $N_1$ and $N_2$ being the number of particles in the halo at $a_1$ and $a_2$ respectively. In these figures, $M_\star$ was defined using top hat (TH) smoothing with $\gamma_f = 4\pi/3$ and the standard $\delta_{c0} = 1.69$, which gave close to the best fit to the FOF(0.2) mass functions in §5. The fits of the analytical to the N-body results in these diagrams are much less sensitive to the adopted value of $\delta_{c0}$ than was the case for the mass functions, and adopting the "best fit" value of $\delta_{c0} = 1.81$ only slightly changes the analytic distributions. In fact, the values of $\delta_{c0}$ which give the best fit for the conditional mass functions are in general somewhat different from those found for the mass functions themselves. The deviations of the first few or last few N-body points from the analytical curves seen in some of the plots seem not to be significant, but depend on the choice of $N_1$ and $\Delta N$. With more conservative cuts, one gets smaller deviations, but then the N-body data covers smaller ranges in $M_1$ and $\Delta M$.

The conditional mass functions, $df/d\ln \Delta M |_{M_1, \Delta a, a_1}$, shown in these figures exhibit a quite complicated dependence on the accreted mass, $\Delta M$, the initial mass, $M_1$, the

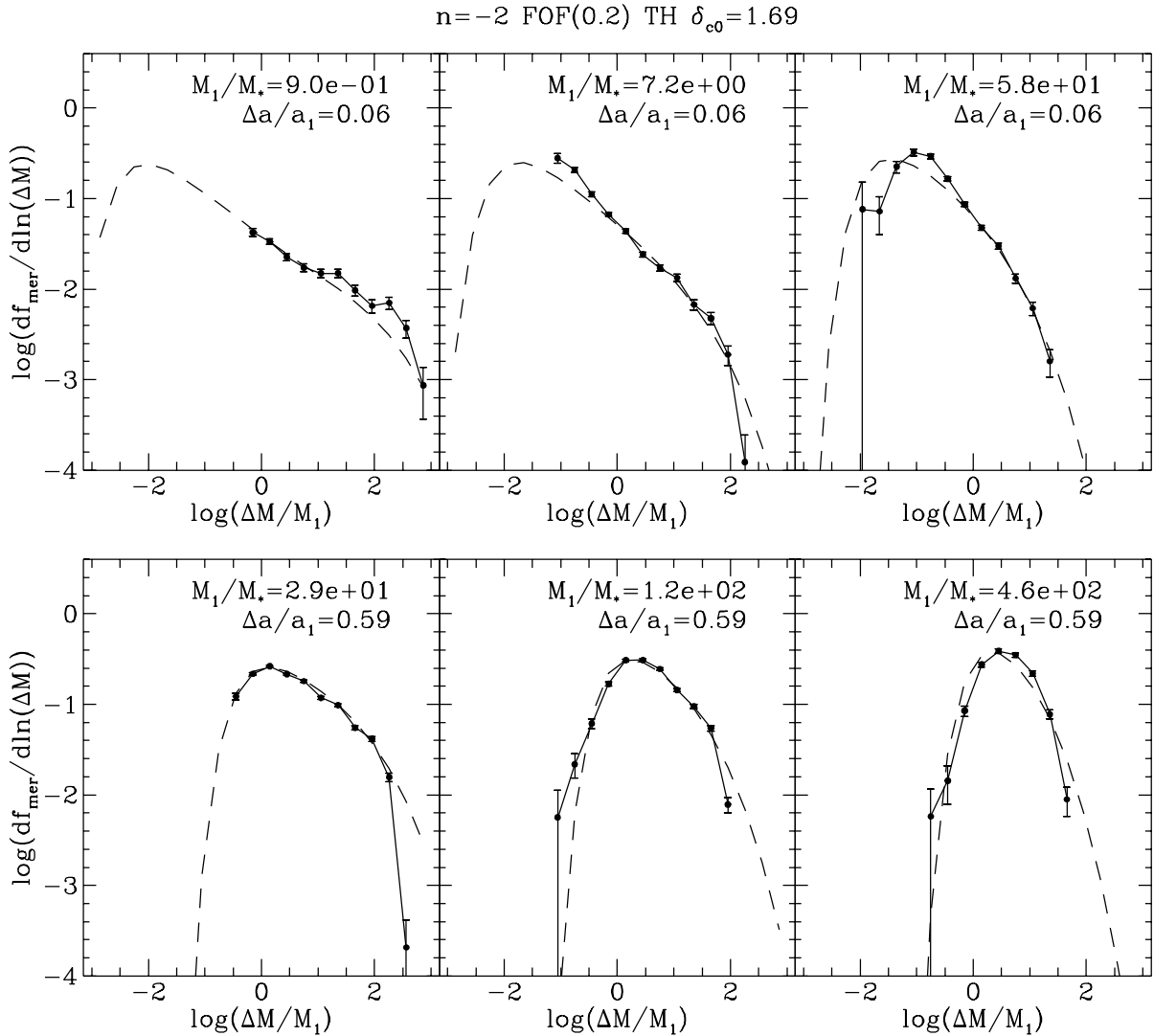

**Figure 4.** A comparison of the analytical conditional mass functions (2.17) (dashed curves) with those of FOF(0.2) groups in the $n = -2$ simulation (solid lines). Top hat (TH) smoothing and the standard $\delta_{c0} = 1.69$ were used to define $M_\star$. We have used only halos with $N_1 > 20$ and $\Delta N > 20$. An indication of the magnitudes of the statistical errors in these estimates are shown by the Poisson error bars, which are calculated as described in the text. The top row of plots show results for an interval $\Delta a/a_1 = 0.06$, which equals that between consecutive outputs of our $n = -2$ simulation, i.e. between which $M_\star$ increases by a factor of $\sqrt{2}$. From left to right these three plots correspond to increasing $M_1/M_\star$ as indicated on each plot. The lower row of plots make the same comparison for a larger time interval, $\Delta a/a_1 = 0.59$, which corresponds to the interval over which $M_\star$ increases by a factor 16.

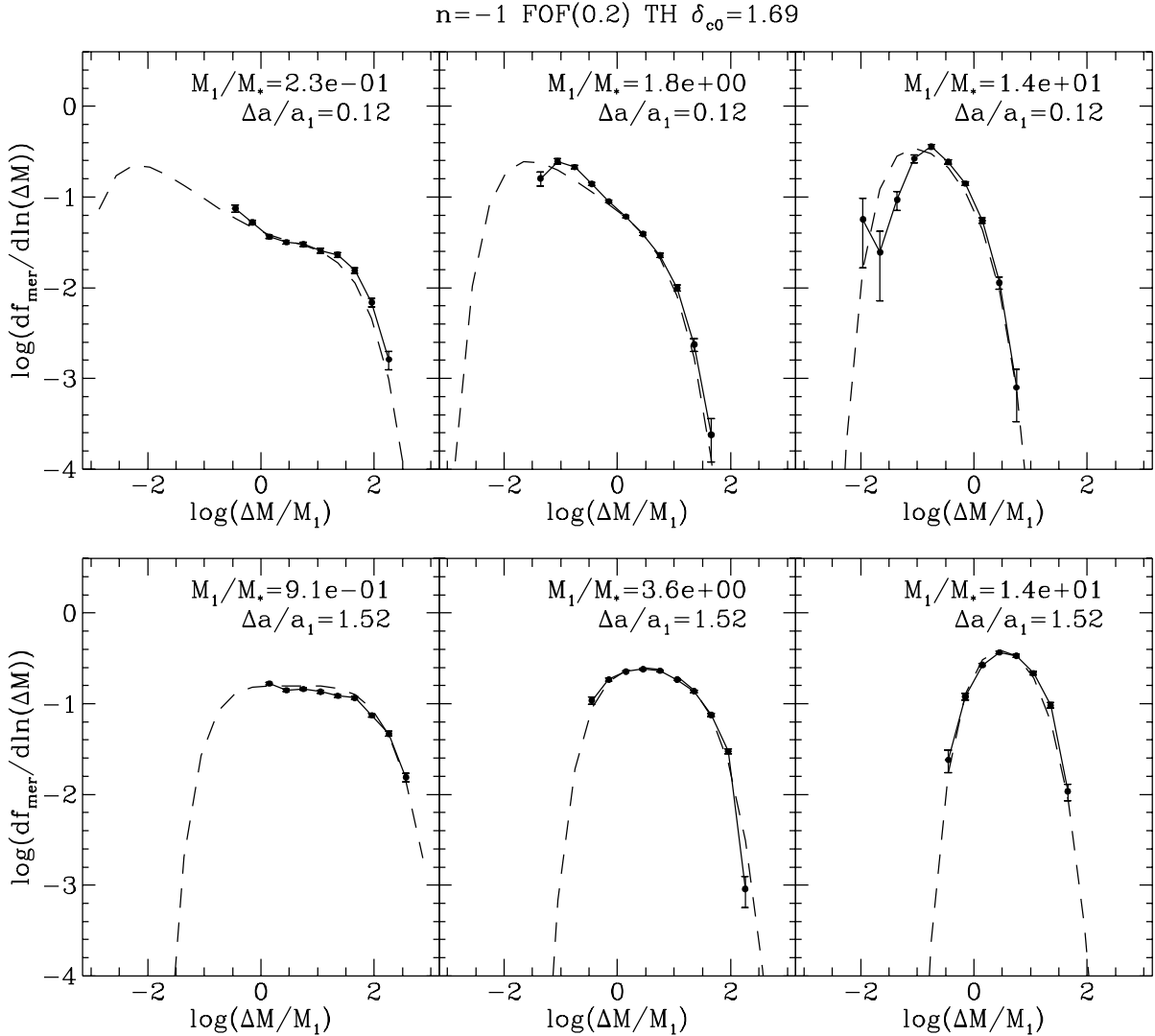

**Figure 5.** Like Figure 4, but for the $n = -1$ model. The values of $\Delta a/a_1$ indicated on the figures again correspond to the intervals over which $M_\star$ increases by factors $\sqrt{2}$ and 16 respectively.

time interval, $\Delta a$, and on the spectral index, $n$, of the initial conditions. The distributions in $\Delta M/M_1$ are asymmetric and vary in both the form and degree of this asymmetry, in their width, and in the location of their peak, when either $M_1/M_\star$, $\Delta a/a_1$, or $n$ are changed. The distributions are broadest and most asymmetric for low values of $M_1/M_\star$. Also for fixed $M_1/M_\star$ they are broader for lower values of $n$. Increasing $\Delta a/a_1$ shifts the peaks of the distributions to higher masses while also altering the asymmetric nature of the low $M_1/M_\star$ distributions. Remarkably all these features are quantitatively reproduced by the analytical expression (2.17). Overall, we find reasonable agreement between the simulations and the analytical distributions over 2-3 decades in $M_1/M_\star$ and $\Delta M/M_1$.

We also investigated the dependence of the conditional mass functions on the choice of group finding algorithm. Using the "best fit" $\delta_{c0}$ values from §5, the analytical distributions fit about as well for FOF(0.3) as for FOF(0.2), while for FOF(0.15) and SO(180), the fits were slightly worse. We show the results for the n=-1 spherical overdensity groups in Figure 7, with $\delta_{c0} = 1.96$. In this case, the analytical distribution actually fits even better if $\delta_{c0} = 1.69$ is chosen.

## 7 FORMATION TIMES

In hierarchical models halos evolve continuously by accreting smaller halos and by merging with comparable and larger halos. Thus there is no clear cut way of defining when a particular halo formed. As a working definition we have adopted the formation time to be the point at which half the mass of a halo is assembled. The distribution $dp_f/dt_f(t_f|M_2, t_2)$, equation (2.19), gives the probability that half the mass of a halo of mass $M_2$ identified at time $t_2$ was assembled at an

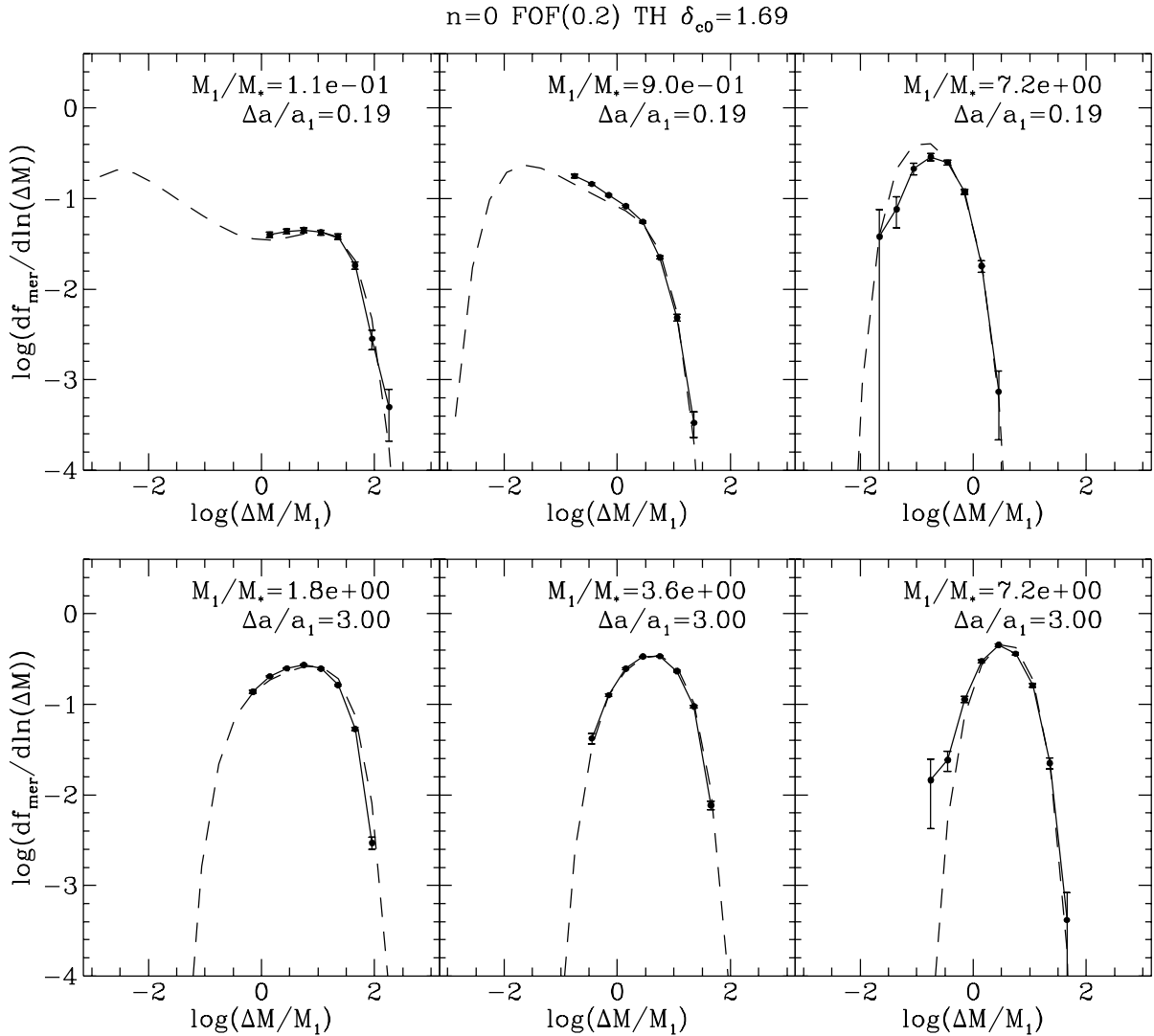

**Figure 6.** Like Figure 4, but for the $n = 0$ model. The values of $\Delta a/a_1$ indicated on the figures again correspond to the intervals over which $M_\star$ increases by factors $\sqrt{2}$ and 16 respectively.

earlier time $t_{\rm f}$. Here we compare this analytical distribution of formation times with estimates made directly from our N-body simulations.

Starting with a set of group catalogues defined using one of the group finding schemes of §5 we proceed as follows. For each group of mass $M$ identified at a particular output epoch at which the expansion factor equals $a_0$, we identify its most massive progenitor at all earlier output times. Progenitors of a group are defined to be all those groups present at an earlier epoch that have the majority (normally greater than 90%) of their particles incorporated into the final group. We then determine at which epoch the mass of this most massive progenitor first becomes larger than $M/2$. This epoch is taken to define the formation time of the halo and defines the expansion factor $a_{\rm f}$ at formation. The histogram of $a_{\rm f}$ values built up for a particular choice of $a_0$ and $M$ defines the formation time distribution $dp_{\rm f}/d\ln a_{\rm f}|_{M,a_0}$. For our scale free simulations, this distribution is a function of only $M/M_\star$ and $a_{\rm f}/a_0$. Hence we can once again combine the histograms from different final output times, $a_0$, to better determine the numerical estimate of $dp_{\rm f}/d\ln a_{\rm f}|_{M/M_\star}$. We choose to use only final epochs separated by a factor 2 in $M_\star$, so that they are approximately independent, and to estimate Poisson errors from the combined number of halos in each bin.

Figures 8, 9 and 10 compare these results for FOF(0.2) groups for a range of $M/M_\star$ with the analytical predictions of equation (2.19), for spectral indices $n = -2, -1$ and 0 respectively. In constructing these plots, we use only halos with $N > 40$ particles at the epoch $a_0$. (Note, if the epoch $a_0$ is identified with the present epoch then the quantity plotted along the x-axis, $\log(a_{\rm f}/a_0) = -\log(1+z_{\rm f})$, where $z_{\rm f}$ is the formation redshift). In each case, there is a clear trend for larger $M/M_\star$ halos to be both younger on average

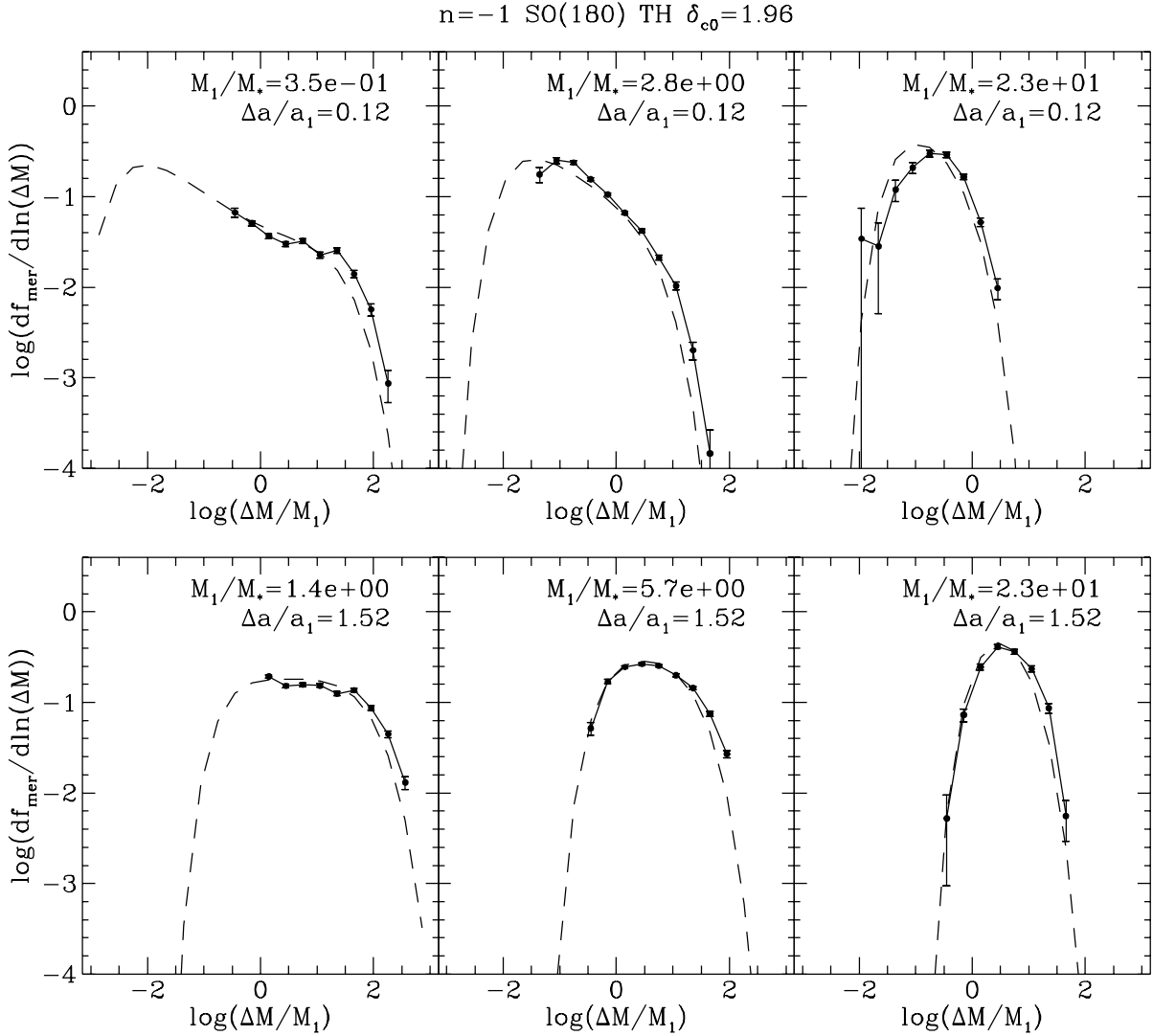

**Figure 7.** Like Figure 5, but for SO(180) groups and with the best fit value of $\delta_{c0} = 1.96$ adopted from Figure 2.

and to have a narrower range of ages. This behaviour, and in fact the precise shape of the distributions, is well reproduced by the analytical distributions given by equation (2.19). We find agreement over 2-3 decades in $M/M_\star$.

We also investigated the formation times of groups identified using the FOF algorithm with differing values of the linking length $b$ and for groups defined by SO with $\kappa = 180$. Once the threshold $\delta_{c0}$ was adjusted to the appropriate "best fit value" found for the mass functions, then the agreement between the analytical and numerical distributions was as good as for the FOF(0.2) groups. In contrast to the fits to the conditional mass functions in §6, the fits to the formation time distributions were appreciably worse for FOF(0.15), FOF(0.3) and SO(180) groups if the value $\delta_{c0} = 1.69$ was used instead of the appropriate "best fit" value. As an example we show the distributions of formation times for the case of the SO(180) groups in Figure 11.

In paper I, we defined a scaled variable

$$\tilde{\omega}_f = \frac{\delta_c(a_f) - \delta_c(a_0)}{(\sigma^2(M/2) - \sigma^2(M))^{1/2}}$$
$$= \frac{(M/M_\star(a_0))^{(n+3)/6}(a_0/a_f - 1)}{(2^{(n+3)/3} - 1)^{1/2}} \qquad (7.1)$$

(the second line is for self-similar models) in terms of which the analytical distribution of formation times $dp_f/d\tilde{\omega}_f$ is independent of $M/M_\star$ and very nearly independent of the spectral index $n$. This last property was demonstrated graphically in Figure 7 of Paper I. Here we transform each of distributions from Figures 8,9 and 10 into distributions $dp_f/d\tilde{\omega}_f$ which we show in Fig. 12. These figures confirm that for the groups found in the N-body simulations there is also very little dependence of these distributions on either mass or spectral index. The short-dashed curves in Fig. 12 are the analytical distributions while the long-dashed curves are the Monte-Carlo distributions taken from Paper I. The Monte Carlo method attempts to follow a single path back through the merger tree leading to a given object, choosing the more massive progenitor each time a halo splits in two, back to when the largest progenitor has fallen below half the final mass. Clearly, the distributions estimated from the N-body simulations are well described by the analytical predictions (equation 2.19), while the Monte Carlo model tends to overestimate halo ages. The problem with the Monte Carlo method appears to lie with an incorrect weighting of the probabilities for the distribution of progenitor masses at each branching point. We are currently working on an improved Monte Carlo procedure which gives results very close to the analytical and N-body ones, and this will be described in a future paper.

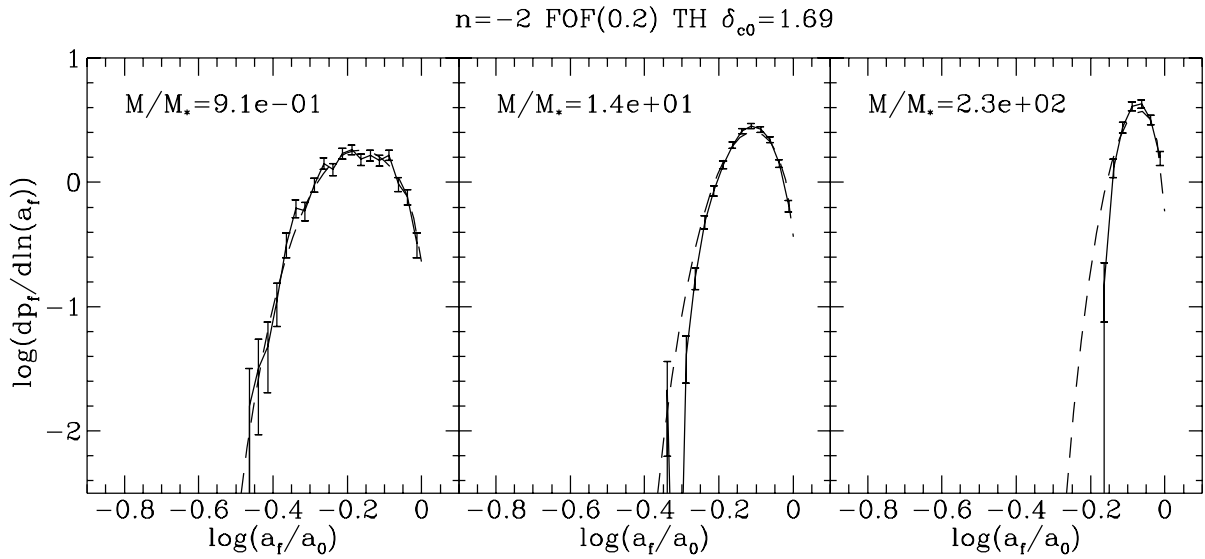

**Figure 8.** Comparison of the distribution of formation epochs derived from the FOF(0.2) groups in the $n = -2$ simulation (solid curves) with the analytical prediction (2.19) (dashed curves), for the various values of $M/M_\star$ indicated. We have assumed $\delta_{co} = 1.69$. For the N-body curves, halos were identified and formation epochs $a_f$ found for a set of identification epochs $a_0$ differing by powers of 2 in $M_\star$, and the results averaged. Only halos with $N > 40$ were used. The error bars represent Poisson errors corresponding to the total number of halos in each bin in $a_f$, after combining the different epochs.

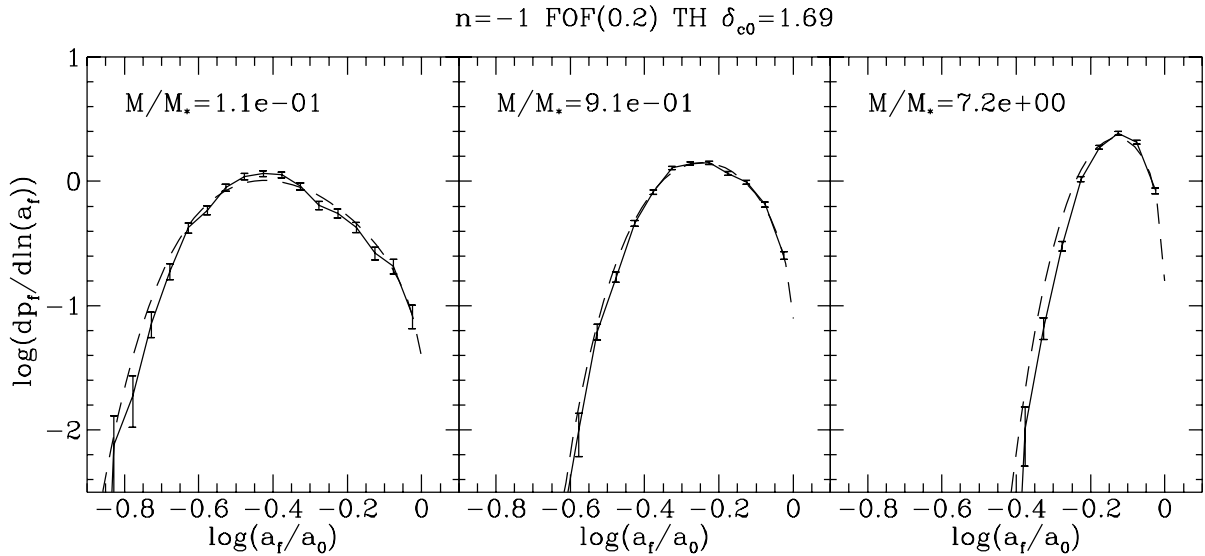

**Figure 9.** Same as Figure 8, but for the $n = -1$ model.

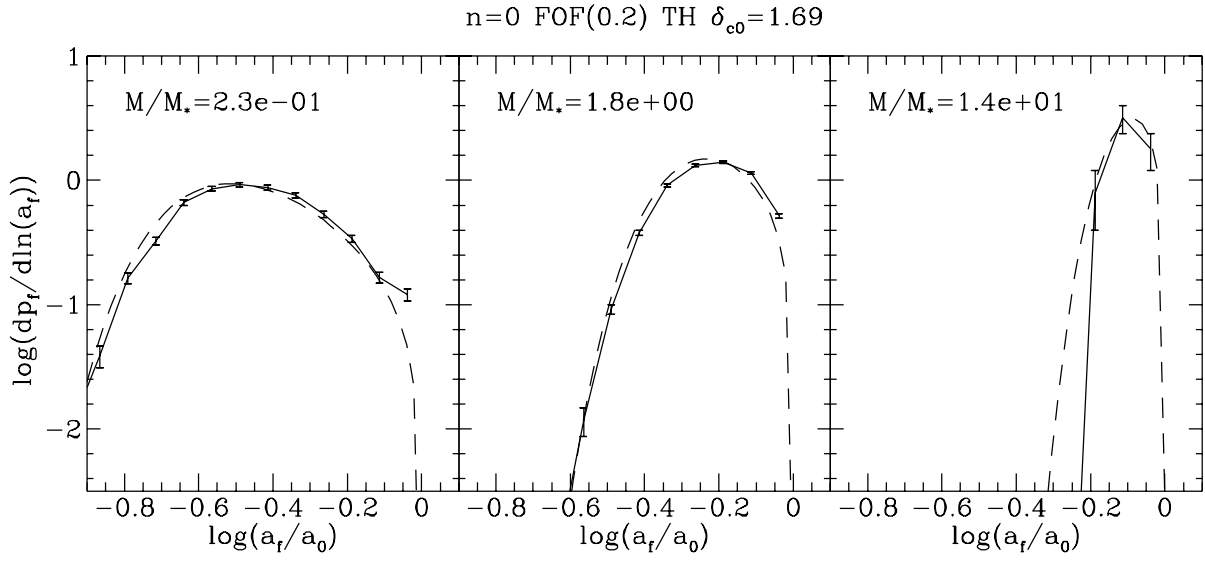

**Figure 10.** Same as Figure 8, but for the $n = 0$ model.

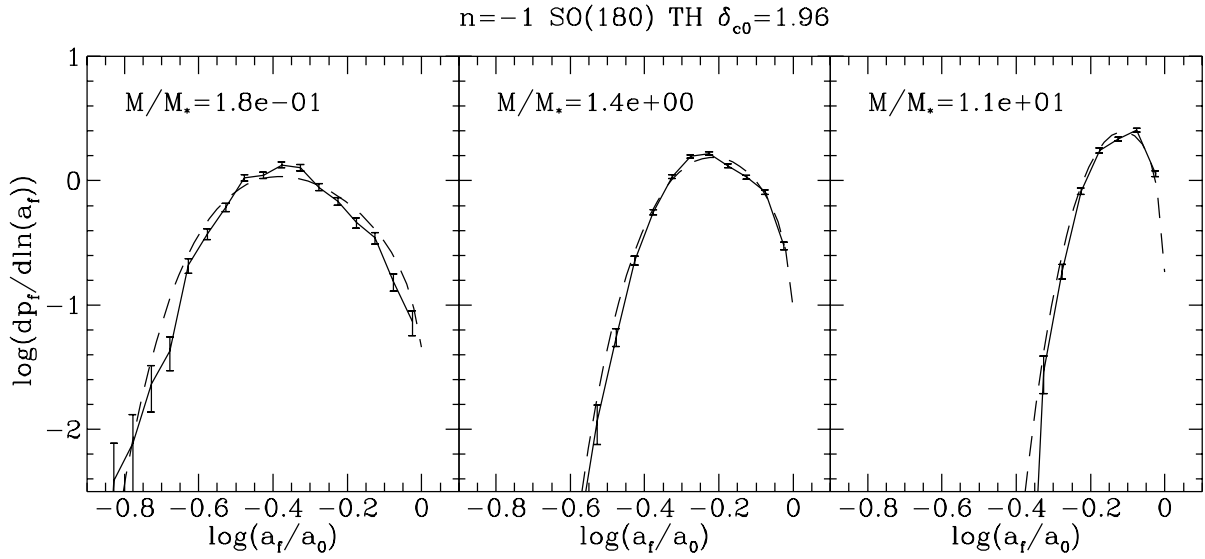

**Figure 11.** Same as Figure 9, but for SO(180) groups in the $n = -1$ model, and with $\delta_{c0} = 1.96$.

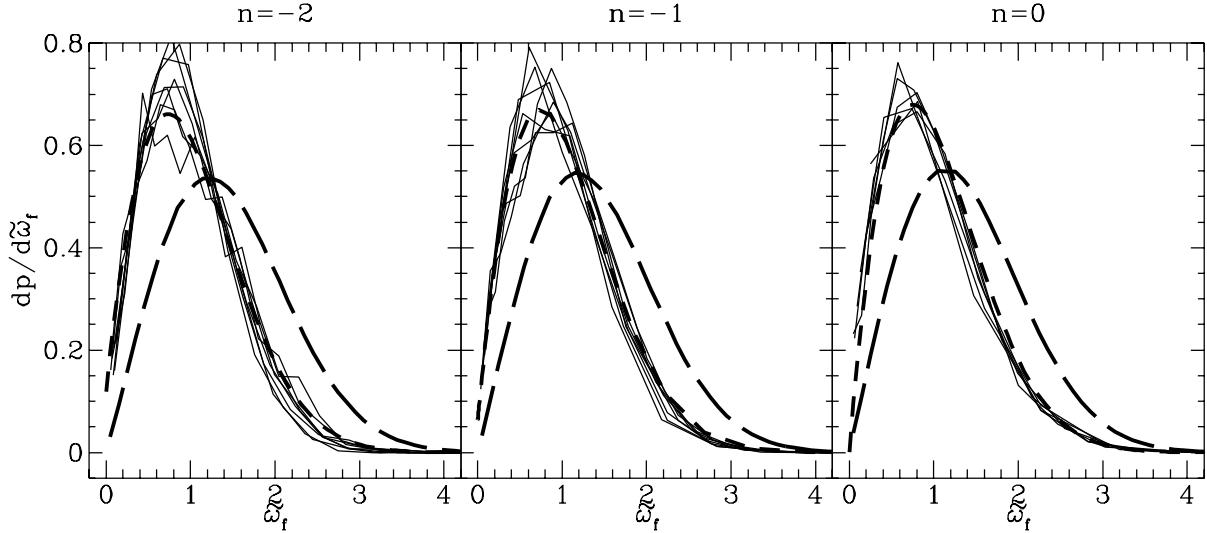

**Figure 12.** Distributions of $dp_f/d\tilde{\omega}_f$ for the N-body simulations compared with the analytical and Monte Carlo predictions from Paper I. The thin solid lines show the N-body results for different values of $M/M_\star$ (averaged over output times as in Figures 8 - 10). The N-body curves plotted are for the mass ranges $M/M_\star = 0.9 - 120, 0.1 - 15, 0.2 - 7$ for $n = -2, -1, 0$ respectively. The short dash and long dash curves are respectively the analytical and Monte Carlo predictions.

## 8 CONCLUSIONS

We have tested the statistical predictions of the Press-Schechter model for the mass function of dark matter halos (Press & Schechter 1974; Bond et al. 1991; Bower 1991), and its extension to halo lifetimes and merger rates (Paper I), against a set of large N-body simulations. The simulations used $128^3$ particles and modelled self-similar clustering with $\Omega = 1$ and initial power spectra $P(k) \propto k^n$, with spectral indices $n = -2, -1, 0$. The comparison reveals that in every respect the analytical formulae produce remarkably good fits to the numerical results. Although tested for self-similar clustering, the analytical formulae are also expected to apply for arbitrary $\Omega$ and more general power spectra, provided that structure grows hierarchically from Gaussian density fluctuations in cold collisionless matter.

Dark matter halos were identified in the simulations using two alternative methods. The first was the standard percolation or "friends of friends" method, which effectively selects objects bounded by surfaces of specified density. We investigated linking lengths $b = 0.15, 0.2$ and $0.3$ in units of the mean interparticle separation, corresponding to mean halo overdensities in the range $\sim 50 - 500$, smaller $b$ corresponding to higher density. The second, "spherical overdensity", method finds spheres of a specified mean density $\kappa$, where we used $\kappa = 180$.

To apply the analytical formulae for mass functions and merging, three choices have to be made: First one must select the form of spatial filter which is applied to the linear density field in order to define $\sigma(R)$, the r.m.s. fluctuation as a function of length scale. Secondly, one must choose a relation between filter radius $R$ and mass $M$ to derive $\sigma(M)$ from $\sigma(R)$. Finally, one must set the critical density threshold for collapse, $\delta_{c0}$. We have investigated top hat, sharp $k$-space and Gaussian filtering, with a mass-radius relation $M = \gamma_f \bar{\rho} R^3$ for some filter-dependent constant, $\gamma_f$. For top hat and sharp $k$-space filtering, the value of $\delta_{c0}$ required to best match the analytical Press-Schechter mass function with the N-body results is independent of the spectral index of the linear density field for $-2 \leq n \leq 0$, when $\gamma_f$ is taken to be the value obtained by integrating the window function over all space (for the top hat, $M$ is just the mass enclosed within radius $R$). For Gaussian filtering, on the other hand, this independence holds only if $\gamma_f$ is taken to be a factor $\sim 2.5$ larger than is given by this integral, contrary to what has been assumed in previous work. When halos are selected using percolation with $b = 0.2$, which selects halos having mean overdensity $\sim 100 - 200$, the best fitting $\delta_{c0}$ values for each of the filters are within 20% of the value $\delta_{c0} = 1.69$ predicted by the analytical model for the collapse of a spherically symmetric overdense region in an $\Omega = 1$ universe. In this case, the Press-Schechter mass function fits the N-body results to an accuracy of $\sim 30\%$. When the group selection method is changed, the best fit $\delta_{c0}$ also changes. The mass functions for the $b = 0.15$ and $\kappa = 180$ groups require larger values of $\delta_{c0}$ (further from $\delta_{c0} = 1.69$) in the Press-Schechter formula to give reasonable fits, compared to the $b = 0.2$ groups, and even then fit somewhat less well.

Overall, there seems little to choose between the different types of filtering, but we have concentrated on the top hat filter in most of our comparisons because this is more standard. For the conventional choices of top hat filtering with $\delta_{c0} = 1.69$, the analytical mass functions differ by less than a factor $1.5 - 2$ from those estimated from the $b = 0.2$ percolation groups in the simulations, over a range of $10^3$ in mass (see Fig. 1). The error is largest for the rare high-mass halos, but is typically $\lesssim 30\%$ for the more numerous halos which contain most of the mass. The error in the mass function can in fact be reduced to $\lesssim 30\%$ overall for this case by increasing $\delta_{c0}$ by $\sim 10\%$. The comparison is limited to the abovementioned range of mass by the dynamic range of our simulations, which span a factor $10^3$ in mass from the smallest resolved halos (containing at least 20 particles) to the most massive. Although still limited in dynamic range, this is a considerable improvement over the comparison made by Efstathiou *et al.* (1988), which utilized simulations containing only 1/64th of the number of particles of our simulations.

With the same choices of filter, threshold, and group-finding, we also find remarkable agreement between the halo merger rates measured from the simulations and the analytical predictions. All the trends seen in the dependence of the merger rate on the masses of the two halos involved and on the time interval considered are reproduced quantitatively by the analytic formula (2.17) (see Figs. 4-6). In fact, equation (2.17) is a reasonable fit to the numerical estimates over the full range of masses, roughly 2-3 decades, that we are able to explore. A similarly impressive agreement, again for a wide range of masses, is seen when one compares the distribution of halo formation times estimated from the simulations with the analytical formula (2.19) (see Figs. 8-10). For the $b = 0.15$, $b = 0.3$ and $\kappa = 180$ groups, the agreement for merger rates and formation times is about as good as for the $b = 0.2$ groups, provided that instead of the standard value $\delta_{c0} = 1.69$, one uses the values of $\delta_{c0}$ which best fit the mass functions in the other formulae too.

We end with a caveat. Despite its success in matching the results of N-body simulations, the Press-Schechter approach from which our formulae are derived falls some way short of being a rigorous analytical model of gravitational instability and non-linear dynamics. It is instead based on the ansatz (Bond *et al.* 1991) that the mass in non-linear objects of mass $M$ can be equated with the mass within regions whose linear theory density perturbation $\delta$ exceeds a threshold value, $\delta_c$, when smoothed on the mass scale $M$, but is below the threshold on all larger scales. In particular, no regard is paid to the shape or size of these regions. Many will enclose less than mass $M$, making it impossible that they form objects of mass $M$ without combining with other nearby material. It is clear that any physically realistic model must take account of the properties of the linear density field across the entirety of each region that collapses to form a non-linear halo. The "peak-patch" analysis of (Bond & Myers 1993a) is the first model that addresses this aspect of the problem. The disadvantage of this more rigorous treatment is that it is very complex and does not lead to analytical formulae for halo mass functions or merger rates. Thus, despite the fundamental flaws in the Press-Schechter approach, the analytical formulae that it yields are extremely valuable if they are an accurate description of the true halo mass functions and merger statistics. This work confirms that these statistical predictions reproduce remarkably well the non-linear hierarchical evolution of dark matter halos in large scale-free cosmological N-body simulations. They therefore provide a valid and extremely useful framework in which to study galaxy and cluster formation in hierarchical models.

## ACKNOWLEDGMENTS

We thank George Efstathiou and Carlos Frenk for providing us with a copy of the $P^3M$ N-body code, and both them and Tom Quinn for much helpful advice on N-body simulations. We also thank Enrique Gaztanaga and the anonymous referee for useful comments which helped improve the paper. CGL is supported by a SERC Advanced Fellowship and SMC by a SERC postdoctoral research assistantship.